\documentclass [aps,prl,twocolumn,showpacs,nofootinbib]{revtex4}
\usepackage{verbatim}
\usepackage{slashed}
\usepackage{epsfig}
\usepackage{graphicx}
\usepackage{amsmath}
\usepackage{mathrsfs}
\usepackage{bm}

\begin{document}


\title{The next-to-leading-order QCD correction to inclusive $J/\psi(\Upsilon)$ production in Z$^0$ decay}



\author{Rong Li and Jian-Xiong Wang}
\affiliation{
Institute of High Energy Physics, Chinese Academy of Sciences, P.O. Box 918(4),
Beijing, 100049, China.\\
Theoretical Physics Center for Science Facilities, CAS, Beijing, 100049, China.
}



\date{\today}

\begin{abstract}

In this paper, we study the $J/\psi(\Upsilon)$ production in Z boson
decay in color-singlet model(CSM). We calculate the next-to-leading-order (NLO) 
QCD correction to $Z \to Quarkonium+Q\bar{Q}$, the dominant 
contribution in the CSM, with the vector and axial-vector parts in $ZQ\bar{Q}$ vertex 
being treated separately. The results show that the vector and
axial-vector parts have the same K factor (the ratio of NLO result to leading-order result) 1.13 
with the renormalization scale $\mu$=2$m_c$ and $m_c=1.5GeV$,
and the K factor falls to 0.918 when applying the Brodsky, Lepage, and Mackenzie(BLM) 
renormalization scale scheme with obtained $\mu_{BLM}=2.28GeV$ and $m_c=1.5$GeV.  
By including the contributions from the next-dominant ones, 
the photon and gluon fragmentation processes, the branching ratio for $Z \to J/\psi_{prompt}+X$ 
is $(7.3 \sim 10.0)\times 10^{-5}$ with the uncertainty consideration for the renormalization scale
and Charm quark mass.  The results are about half of the central value of the experimental 
measurement 2.1$\times10^{-4}$. Furthermore, the $J/\psi$ energy distribution in our calculation 
is not well consistent with the experimental data. Therefore, even at QCD NLO, the contribution to 
$Z \to J/\psi_{prompt}+X$ from the CSM can not fully account for the experimental measurement. 
And there should be contributions from other mechanisms, such as the color-octet(COM) contributions. 
We define $R_{c\bar{c}}=\frac{\Gamma(Z \to J/\psi  c\bar{c}X)}{\Gamma(Z \to J/\psi  X)}$
and obtain $R_{cc}=0.84$ for only CSM contribution and $R_{cc}=0.49$ for COM and CSM
contributions together. Then $R_{cc}$ measurement could be used to clarify
the COM contributions.
\end{abstract}

\pacs{12.38.Bx, 13.38.Dg, 14.40.Pq, 12.39.Jh}

\maketitle


\section{I. Introduction}

Heavy Quarkonium is an ideal system being used to study the
perturbative and non-perturbative aspects of QCD. Firstly, the heavy
quark mass sets a large scale for perturbative calculation.
Secondly, the dileptonic decay of heavy quarkonium makes the
identification and measurement efficient. In 1995,
the non-relativistic QCD(NRQCD), a rigorous effective theory in
describing the production and decay of heavy quarkonium,
was proposed~\cite{Bodwin:1994jh}, and it makes the
color-singlet model(CSM)~\cite{Einhorn:1975ua} be its leading-order
approximation in $v$ (the velocity between heavy quark and
anti-quark in the meson rest frame). More details on NRQCD and heavy
quarkonium physics can be found in reference~\cite{Brambilla:2004wf}.

In recent years, there are many works on the next-to-leading-order(NLO)
QCD correction for heavy quarkonium productions. To explain the experimental 
measurement~\cite{Abe:2001za,Aubert:2005tj} of $J/\psi$ production at the B factories, a
series of calculations~\cite{Zhang:2005cha,Gong:2009ng} have been performed and 
revealed that the NLO QCD corrections can change the
leading-order(LO) theoretical predictions considerably and the NLO
results in CSM give the main contribution to the related processes. 
Together with the relativistic correction~\cite{Bodwin:2006ke}, it seems that all the experimental 
data for $J/\psi$ production at the B factories could be understood.
For $J/\psi$ production in the hadron colliders, there are obviously progress in the 
theoretical calculation. The NLO QCD correction to the CSM
processes\cite{Campbell:2007ws,Gong:2008sn} greatly enhanced the $p_t$ (transverse momentum of $J/\psi$) 
distribution of $J/\psi$ production at large $p_t$ region, and the $p_t$ distribution 
of $J/\psi$ polarization is drastically changed from mostly transverse polarization at 
LO into mostly longitudinal polarization at NLO~\cite{Gong:2008sn}. 
It is found that the NLO QCD correction to $J/\psi$ production for color-octet (COM) parts
is quite small, about 10 percent~\cite{Gong:2008ft}. Even including all these progresses, we still 
can not obtain an satisfactory explanation on both the $p_t$ distribution of the production and 
polarization for $J/\psi$ hadroproduction. The partial next-to-next-to-leading-order(NNLO) calculations
for $\Upsilon$ and $J/\psi$ hadroproduction show that the uncertainty form QCD higher order
correction~\cite{Artoisenet:2008fc} is much bigger, but still can not cover the $J/\psi$ or $\Upsilon$ 
polarization measurement. Recent studies reveal that the NLO QCD correction also plays an 
important role on $J/\psi$ production at RHIC~\cite{Brodsky:2009cf} and the 
hadroproduction of $\chi_c$~\cite{Ma:2010vd}. 
The $J/\psi$ photoproduction once was considered as 
an positive example with the $p_t$ and $z$ distribution well described by the NLO calculations 
in CSM~\cite{Kramer:1995nb}. But either the $p_t$ distribution of the production or 
polarization for $J/\psi$ can not be well described by the recent NLO calculations 
in CSM \cite{Artoisenet:2009xh}. It seems that the complete calculation at NLO in COM ~\cite{Butenschoen:2009zy} 
can account for the experimental measurements on the $p_t$ distribution. 
But the complete calculation on $J/\psi$ polarization at NLO in COM is 
a real challenge.

With both the successful and unsuccessful aspects for 
theoretical progress in heavy quarkonium production, it is worthwhile to 
study more cases in detail. Such as $J/\psi$ production associated with 
photon~\cite{Li:2008ym},
QED contributions in $J/\psi$ hadroproduction~\cite{He:2009cq},  
inclusive $J/\psi$ production from $\Upsilon$ decay~\cite{He:2009by} and $J/\psi$ production from $Z$ decay.
Heavy quarkonium production in Z decay has been widely studied
in the CSM and COM at LO~\cite{Guberina:1980dc,Cheung:1995ka,Baek:1996kq,Fleming:1993fq}, 
and the measurement at the LEP by L3 Collaboration
gives the branching ratio as~\cite{Acciarri:1998iy}
\begin{flalign}
&Br(Z \to
J/\psi_{prompt}+X) \nonumber \\
&=(2.1\pm0.6(stat.)\pm0.4(sys.)^{+0.4}_{-0.2}(theo.))\times10^{-4},\label{exj} \\
\nonumber \\
&Br(Z \to \Upsilon(1S)+X)<4.4 \times 10^{-5}.
\end{flalign}
Theoretical investigation on this process indicates that even the
dominant channel $Z \to J/\psi+c+\bar{c}$ in all the CSM ones at LO only gives the 1/3 prediction to the
total branching ratio of the experimental measurement. Including the
contribution of gluon fragmentation process in the COM can enhance the
theoretical results about 3 times~\cite{Cheung:1995ka}. 
This once is an evidence for the effect of the COM.  
It also have been studied in color-evaporation model(CEM) in reference~\cite{Gregores:1996ek} and
obtained consistent results with the experimental data. But the CEM always gives unpolarized 
$J/\psi$ in conflict with experimental measurements. By resuming the large logarithm from the
large difference of $J/\psi$ and Z mass, the COM prediction
on $J/\psi$ energy distribution($d\Gamma/dz$ with $z$=2$E_{J/\psi}/M_z$) is roughly consist with the
data~\cite{Boyd:1998km}. Considering the larger impact of the NLO QCD corrections to the production 
of heavy quarkonium, it is necessary to investigate the NLO QCD correction to $Z \to
J/\psi+c\bar{c}+X$.
In this paper, we calculate the NLO QCD correction to $Z \to
J/\psi+c\bar{c}+X$, and also include the contributions from the gluon and photon
fragmentation processes as well with only the CSM in consideration. The study could
provide more insight to the effect of color-octet mechanism and put more constrains on
the value of color-octet matrix elements.

This paper is organized as follows. In Sec. II, we study the NLO QCD correction to the 
heavy quark association process with different schemes on the choice of renormalization 
scale. In Sec. III, we investigated dominant fragmentation processes and give the total 
results on prompt $J/\psi$ production in Z decay. In Sec. IV, the summary and conclusion are presented.

\section{II. The heavy quark association process}
\begin{figure}
\center{
\includegraphics[scale=0.65]{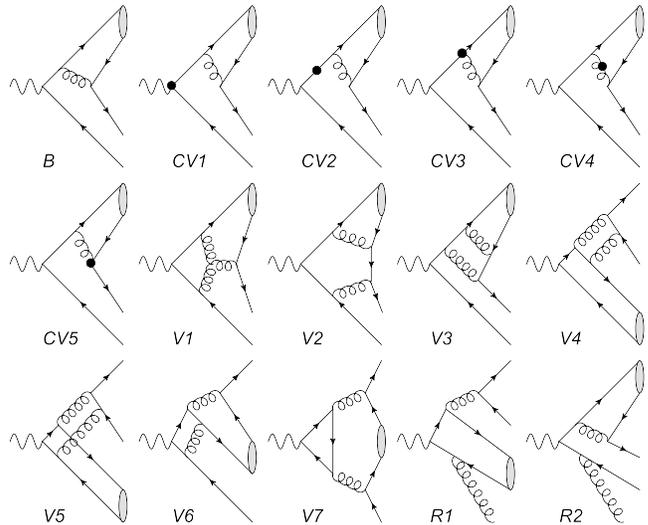}
\caption{\label{fig:feynman}Typical Feynman diagrams for
$J/\psi+c\bar{c}+X$ production in Z decay. B is the Born diagram.
CV1$\sim$CV5 represent the counterterm diagrams and 
corresponding loop diagrams. V1$\sim$V7 represent the box and the
anomalous triangle diagrams. R1 and R2 are the real parts.}}
\end{figure}

\begin{figure}
\center{
\includegraphics[scale=0.65]{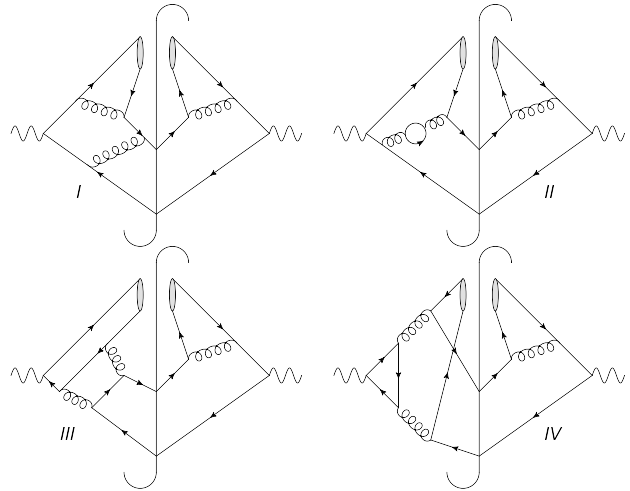}
\caption{\label{fig:cutdiagram}The typical amplitude squared diagrams for
$J/\psi+c\bar{c}+X$ production in Z decay.}}
\end{figure}

For the calculation on $Z \to J/\psi + c \bar{c} +X $ at NLO, there are virtual and real \
correction parts as
\begin{eqnarray}\label{singlet}
Z \to J/\psi +c+\bar{c} \\
Z \to J/\psi +c+\bar{c}+g.
\end{eqnarray}
There are vector and axial-vector parts in the coupling of Z boson to fermions,
but the interference between them does not contribute in our calculation. Therefore, we 
study them separately. There are 4 Feynman diagrams for both parts at LO, 80 for the
vector part and 76 for the axial-vector part at NLO. The typical diagrams
are presented in Fig.~\ref{fig:feynman}.
The dimensional regularization is used to regulate the 
the ultraviolet (UV) and infrared (IR) divergence,  and the Coulomb singularity is regulated by introducing
a small relative velocity between quark pair in the quarkonium and absorbed into 
the wave function of quarkonium. In calculating the axial-vector part, we have to face 
the $\gamma_5$ problem.  The structure of all the amplitude squared diagrams 
could be classified into four cases shown in Fig.~\ref{fig:cutdiagram}.

Case 1. There are only one fermion-loop and two $\gamma_5$ matrices appear
in it. Then $\gamma_5$s can be moved together and give
an identity matrix by $\gamma^2_5=1$.

Case 2. There are two fermion-loops and the two $\gamma_5$ matrices appear
in one of them. It is the same as case 1.

Case 3. There are two fermion-loops. Each of them has a $\gamma_5$.
Because there are no UV and IR divergences in the loops, the
dimension can be set as 4 safely.

Case 4. The only special case are the two triangle anomalous diagrams. In
this case we use the scheme described in reference
\cite{Korner:1991sx} to handle it, which fixes the starting point to
write down all the amplitude and abandon the cyclicity in
calculating the trace of the fermion-loop with odd number of
$\gamma_5$. This two triangle anomalous diagrams will not contribute at all
according to Yang's theorem \cite{Yang:1950rg} when the two gluon lines are on mass-shell,
but will contribute in our case since the two connected gluons are off mass-shell.

The on-mass-shell (OS) scheme is used to define the renormalization
constants $Z_m$, $Z_2$ and $Z_3$, which correspond to charm quark
mass $m_c$, charm field $\psi_c$, and gluon field $A^a_\mu$ while
$Z_g$ for the QCD gauge coupling $\alpha_s$ is defined in the
modified-minimal-subtraction($\overline{\mathrm{MS}}$) scheme:
\begin{eqnarray}
\delta Z_m^{\mathrm{OS}}&=&-3C_F\dfrac{\alpha_s}{4\pi}[\dfrac{1}{\epsilon_{UV}} -\gamma_E +\ln\dfrac{4\pi \mu^2}{m_c^2} +\frac{4}{3}] , \nonumber \\
\delta Z_2^{\mathrm{OS}}&=&-C_F\dfrac{\alpha_s}{4\pi}\biggl[\dfrac{1}{\epsilon_{UV}} +\dfrac{2}{\epsilon_{IR}} -3\gamma_E +3\ln\dfrac{4\pi \mu^2}{m_c^2} +4 \biggr] , \nonumber \\
\delta
Z_3^{\mathrm{OS}}&=&\dfrac{\alpha_s}{4\pi}\biggl[(\beta'_0-2C_A)(\dfrac{1}{\epsilon_{UV}}
-\dfrac{1}{\epsilon_{IR}}) \nonumber \\&&
-\dfrac{4}{3}T_F(\dfrac{1}{\epsilon_{UV}} -\gamma_E +\ln\dfrac{4\pi \mu^2}{m_c^2}) \biggr] , \nonumber \\
\delta
Z_g^{\overline{\mathrm{MS}}}&=&-\dfrac{\beta_0}{2}\dfrac{\alpha_s}{4\pi}[\dfrac{1}{\epsilon_{UV}}
-\gamma_E +\ln(4\pi)] .
\end{eqnarray}
where $\mu$ is the renormalization scale, $\gamma_E$ is Euler's
constant, $\beta_0=\frac{11}{3}C_A-\frac{4}{3}T_Fn_f$ is the
one-loop coefficient of the QCD beta function and $n_f$ is the
number of active quark flavors. There are three massless light
quarks $u, d, s$, and heavy quark $c$, so $n_f$=4. In $SU(3)_c$,
color factors are given by $T_F=\frac{1}{2}, C_F=\frac{4}{3},
C_A=3$. And
$\beta'_0\equiv\beta_0+(4/3)T_F=(11/3)C_A-(4/3)T_Fn_{lf}$ where
$n_{lf}\equiv n_f-1=3$ is the number of light quarks flavors.
Actually in the NLO total amplitude level, the terms proportion to
$\delta {Z_3}^{\mathrm{OS}}$  cancel each other, thus the result is
independent of renormalization scheme of the gluon field.
The above renormalization scheme and constant are similar 
to those in reference \cite{Gong:2007db}.
The bottom quark should be considered for 
the calculation of $\Upsilon$ production.

We use the Feynman Diagram Calculation(FDC)
package~\cite{Wang:2004du} to generate Feynman diagram and amplitude,
to do the tensor reduction and scalar integration, and to give the FORTRAN 
code for numerical calculation finally. Because there are some large numbers 
generated in the program, the quadruple precision FORTRAN source is used.

\begin{figure}
\center{
\includegraphics*[scale=0.4]{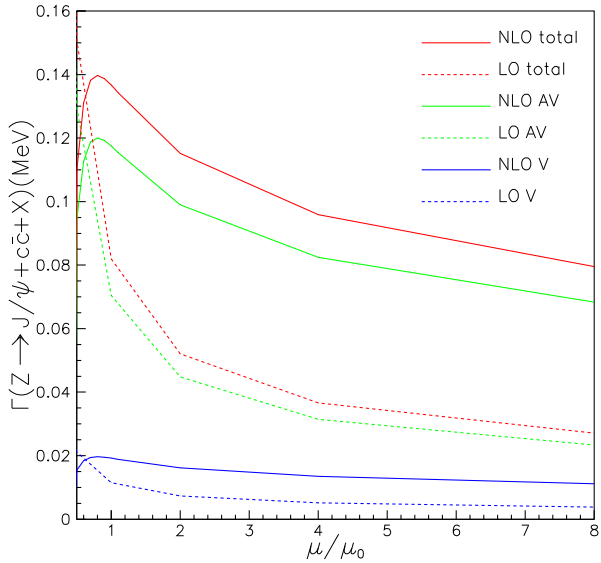}
\caption {\label{fig:jscale1}The $\mu$ dependence of the partial decay width for
$Z\rightarrow J/\psi c\bar{c}+X$ with $m_c=1.5$GeV and $\mu_0=2m_c$. Here the LO results
are calculated with the wave function at origin at LO and the $\alpha_s$ are fixed by one-loop 
running, and these choices are also applied for the LO plots in Fig.~\ref{fig:uscale1},
~\ref{fig:jdis} and \ref{fig:udis}. In all the figures, V presents the 
vector part result, A-V presents the axial-vector part and total 
presents the sum of these two parts. }}
\end{figure}

\begin{figure}
\center{
\includegraphics*[scale=0.4]{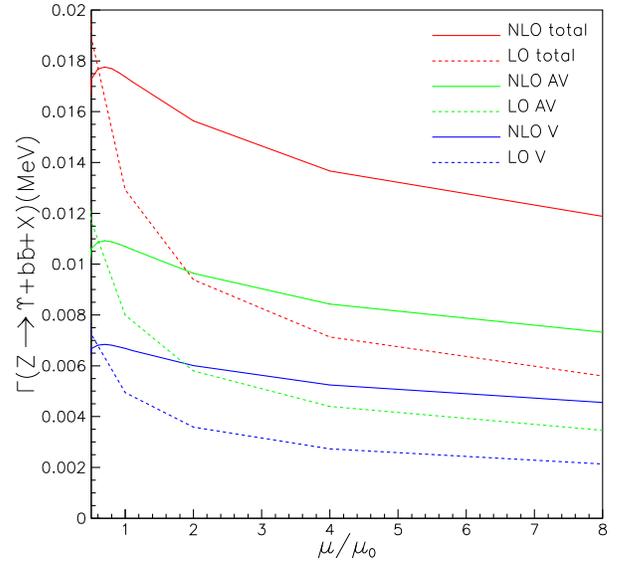}
\caption {\label{fig:uscale1}The $\mu$ dependence of the partial decay width for
$Z\rightarrow \Upsilon b\bar{b}+X$ with $m_b=4.75$GeV and $\mu_0=2M_b$. }}
\end{figure}

\begin{figure}
\center{
\includegraphics*[scale=0.4]{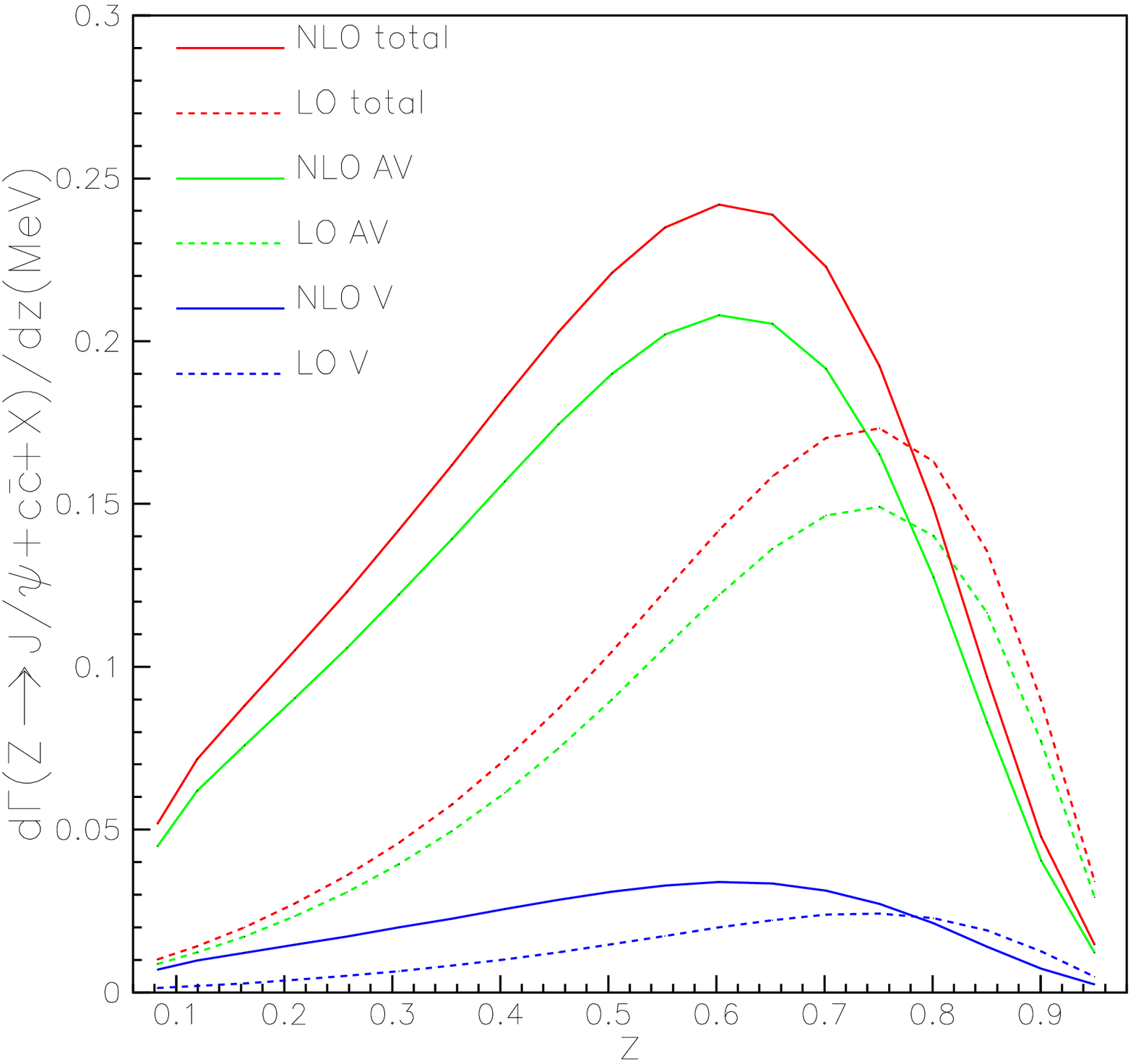}
\caption {\label{fig:jdis}The $J/\psi$ energy distribution in $Z\rightarrow
J/\psi c\bar{c}$ with $m_c=1.5$GeV and $\mu=2M_c$.
}}
\end{figure}

\begin{figure}
\center{
\includegraphics*[scale=0.4]{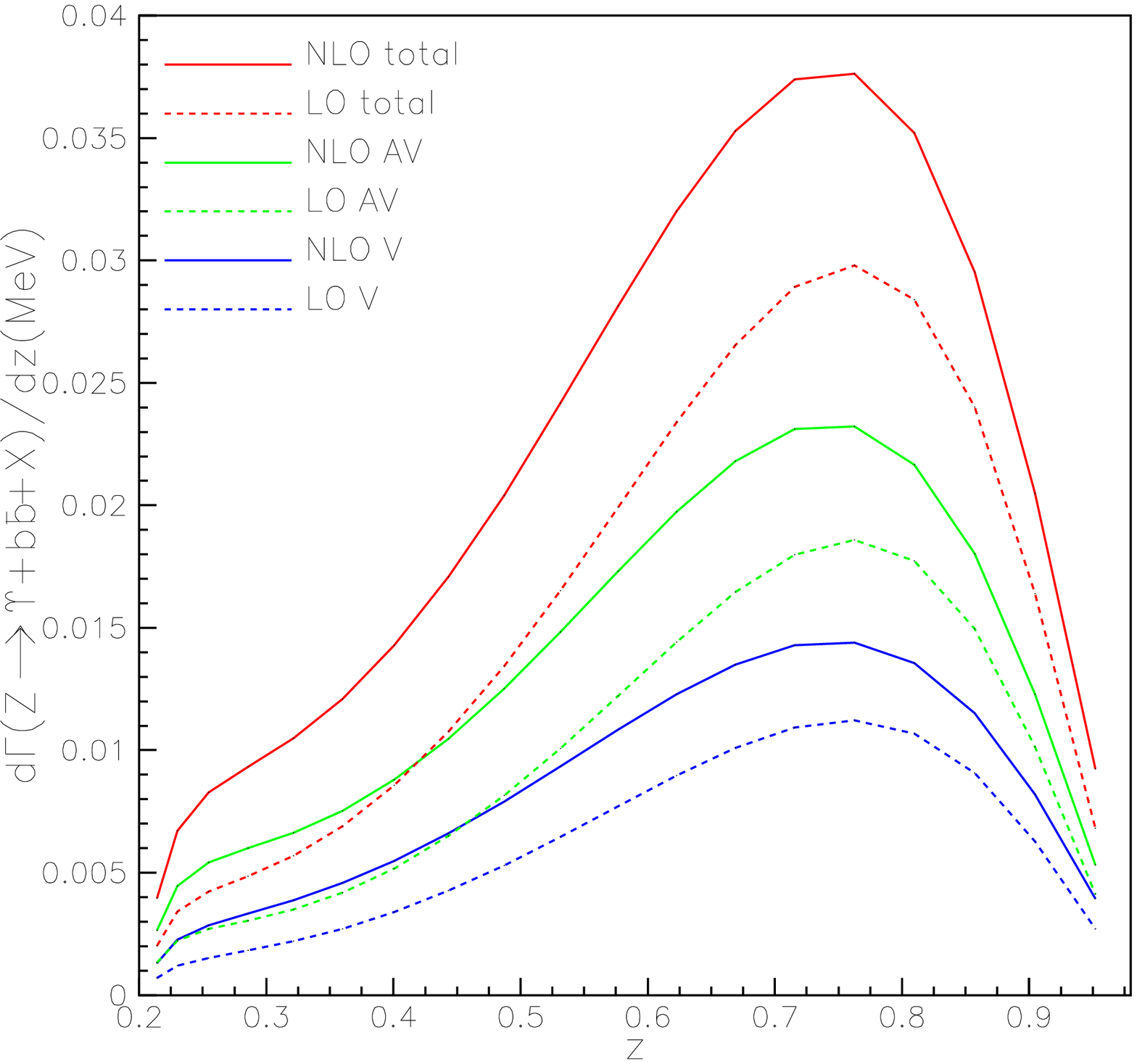}
\caption {\label{fig:udis}The $\Upsilon$ energy distribution in $Z\rightarrow
\Upsilon b\bar{b}$ with $m_b=4.75$GeV and $\mu=2M_b$.
}}
\end{figure}

The leptonic width of $J/\psi(\Upsilon)$ is used to extract their
wave functions at origin $R_s^{J/\psi(\Upsilon)}$, which is
\begin{eqnarray}
\Gamma_{ee}=(1-\frac{16\alpha_s}{3\pi})\frac{4\alpha^2
e_{c(b)}^2}{M_{J/\psi(\Upsilon)}^2}|R_s^{J/\psi(\Upsilon)}|^2.\nonumber
\end{eqnarray}
Here the values of the parameters are chosen as
$\Gamma_{ee}^{J/\psi}=5.55$keV,
$\Gamma_{ee}^{\Upsilon}=1.34$keV~\cite{Amsler:2008zz},
$\alpha$=1/137 and $\alpha_s=\alpha_s^{2loop}(2m_Q)$. The one-loop
and two-loop running program of CTEQ6 are used to fix the LO and NLO
values of $\alpha_s$. The LO wave functions of heavy quarkonium are
used to obtain the LO results in
Fig.~\ref{fig:jscale1},~\ref{fig:uscale1},~\ref{fig:jdis} and
~\ref{fig:udis}. In the following calculation, $\alpha=1/128$ is used, 
and the central value of heavy quark mass is chosen as
$m_c$=1.5GeV ($m_b$=4.75GeV). We also use
$m_c$=1.4,1.6 GeV ($m_b$=4.65, 4.85 GeV) for uncertainty estimate. The default choice of
renormalization scale is 2$m_c$(2$m_b$) for $J/\psi$ ($\Upsilon$).

The LO and NLO partial decay width of $Z \to J/\psi + c \bar{c} +X$
are presented in Table ~\ref{table:resultc}. The difference between 
our LO results and the other LO results in literature is mainly due to the different choice of 
the wave functions at origin. The QCD correction
enhance the partial decay width about 13$\%$ for both the vector
part and the axial-vector part when the same wave function at origin is used. 
This may provide a hint that the picture of heavy quark fragmentation into quarkonium works 
at these energy scale at NLO. It can also be seen that the K factors are 
insensitive to the variance of the quark mass. 
\begin{table*}[htbp]
\begin{center}
\begin{tabular}{|c|c|c|c|c|c|c|c|c|}
\hline\hline
$m_c$(GeV)&$\alpha_s(\mu)$&$\Gamma_V^{(0)}$(keV)&$\Gamma_V^{(1)}$(keV)&$\Gamma_V^{(1)}$/$\Gamma_V^{(0)}$&$\Gamma_{AV}^{(0)}$(keV)&$\Gamma_{AV}^{(1)}$(keV)&$\Gamma_{AV}^{(1)}$/$\Gamma_{AV}^{(0)}$&$\Gamma_{tot}^{(1)}$/$\Gamma_{tot}^{(0)}$\\
\hline
1.4&0.266&19.6&22.2&1.13&120&136&1.13&1.13 \\
\hline
1.5&0.259&16.9&19.1&1.13&103&117&1.13&1.13 \\
\hline
1.6&0.252&14.8&16.7&1.13&90.0&102&1.13&1.13 \\
\hline\hline
\end{tabular}
\caption{\label{table:resultc}The partial decay width for $J/\psi$ with the renormalization scale $\mu=2m_c$
and different charm quark mass $m_c$.}
\end{center}
\end{table*}
For the $\Upsilon$ production the similar results are presented in
Table ~\ref{table:resultu}. And it is easy to find that there is very small difference in K factors 
between the vector part and the axial-vector part. It could be thought as that the large 
bottom quark mass makes the fragmentation picture less effective than that in the $J/\psi$ 
production process.

\begin{table*}[htbp]
\begin{center}
\begin{tabular}{|c|c|c|c|c|c|c|c|c|}
\hline\hline
$m_b$(GeV)&$\alpha_s(\mu)$&$\Gamma_V^{(0)}$(keV)&$\Gamma_V^{(1)}$(keV)&$\Gamma_V^{(1)}$/$\Gamma_V^{(0)}$&$\Gamma_{AV}^{(0)}$(keV)&$\Gamma_{AV}^{(1)}$(keV)&$\Gamma_{AV}^{(1)}$/$\Gamma_{AV}^{(0)}$&$\Gamma_{tot}^{(1)}$/$\Gamma_{tot}^{(0)}$\\
\hline
4.65&0.184&5.50&6.88&1.24&8.95&11.1&1.25&1.24 \\
\hline
4.75&0.183&5.33&6.68&1.24&8.61&10.7&1.25&1.25 \\
\hline
4.85&0.182&5.17&6.49&1.24&8.29&10.3&1.26&1.25 \\
\hline\hline
\end{tabular}
\caption{The partial decay width for $\Upsilon$ with the renormalization scale $\mu=2m_b$ 
and different bottom quark mass $m_b$.} \label{table:resultu}
\end{center}
\end{table*}

The renormalization scale dependence of the partial decay widths for $J/\psi$ and
$\Upsilon$ are shown in Fig.~\ref{fig:jscale1} and
Fig.~\ref{fig:uscale1}. The QCD correction improve the scale
dependence in small $\mu$ region and there are similar behavior for
LO and NLO results in other region.
In Fig.~\ref{fig:jdis} and ~\ref{fig:udis}, the energy
distribution of $J/\psi$ and $\Upsilon$ are shown with $z$ defined
as $2E_{J/\psi(\Upsilon)}/M_Z$. The NLO QCD correction shifts the
maximum point of $J/\psi$ energy distribution from the large $z$
region to the middle $z$ region. But for $\Upsilon$, the shifts is not
manifest.

To study the uncertainty from different choices of the renormalization scale, 
in addition to our default choice $2m_c(m_b)$ for $J/\psi(\Upsilon)$ in the calculation, 
we use other two schemes to fix the renormalization scale. At first, 
the decay width could be expressed as
\begin{eqnarray}\label{total}
\Gamma^{NLO}(\mu)=\Gamma^{LO}(\mu)[1+\frac{\alpha_s(\mu)}{\pi}(A+\beta^\prime_0ln\frac{\mu}{2m_Q}+Bn_f)].
\end{eqnarray}
Here the LO results depend on the renormalization scale through the
running of the coupling constant. A and B are independent of the
scale and $\beta^\prime_0=11-2n_{lf}/3$. We extract the parameters in Eq.~\ref{total} 
and present them in Table~\ref{table:para}.

\begin{table}[htbp]
\begin{center}
\begin{tabular}{|c|c|c|c|}
\hline\hline
$m_c$(GeV)&$\Gamma_{J/\psi}^{LO}$(keV)&A&B\\
\hline
1.40&176&2.08&-0.178 \\
\hline
1.50&151&2.12&-0.182 \\
\hline
1.60&131&2.16&-0.186 \\
\hline
$m_b$(GeV)&$\Gamma_{\Upsilon}^{LO}$(keV)& & \\
\hline
4.65&17.8&4.97&-0.273 \\
\hline
4.75&17.2&5.05&-0.275 \\
\hline
4.85&16.6&5.12&-0.278 \\
\hline\hline
\end{tabular}
\caption{The extracted parameters for Eq.~\ref{total}} \label{table:para}
\end{center}
\end{table}

Scheme I: From Fig.~\ref{fig:jscale1} and Fig.~\ref{fig:uscale1}, it can
be seen that there are the $\mu$ points where the partial decay widths reach
their maximum values. By using the Eq.~\ref{total}, the values of $\mu$
and partial decay widths can be obtained and presented in
Table~\ref{table:schemei}.
\begin{table}[htbp]
\begin{center}
\begin{tabular}{|c|c|c|c|}
\hline\hline
$m_c$(GeV)&$\mu$(GeV)&$\Gamma_{J/\psi}^{NLO}$(keV)\\
\hline
1.40&2.26&162 \\
\hline
1.50&2.42&139 \\
\hline
1.60&2.58&120 \\
\hline
$m_b$(GeV)& &$\Gamma_{\Upsilon}^{NLO}$(keV)\\
\hline
4.65&6.48&18.4 \\
\hline
4.75&6.57&17.8 \\
\hline
4.85&6.66&17.2 \\
\hline\hline
\end{tabular}
\caption{The maximum partial decay width for $Z \to J/\psi(\Upsilon)+c\bar{c}(b\bar{b})+X$ in the scheme I.}
\label{table:schemei}
\end{center}
\end{table}

Scheme II: In Brodsky, Lepage, and Mackenzie(BLM)
scheme~\cite{Brodsky:1982gc}, the $n_{lf}$(light quark flavor)
dependence of the QCD correction is absorbed into the running of
$\alpha_s$ by shifting the renormalization scale. An improved result
on process $e^+e^-\rightarrow J/\psi c\bar{c}$ has been obtained in reference
\cite{Gong:2009ng}. So we also try this scheme in our calculation and the results are 
presented in Table \ref{table:resultJ} and \ref{table:resultU}.
\begin{table}[htbp]
\begin{center}
\begin{tabular}{|c|c|c|c|c|c|}
\hline\hline
$m_c$(GeV) & $\mu^*$(GeV) & $\alpha_s(\mu^*)$ & $\Gamma^{(0)}$(keV) & $\Gamma^{(1)}$(keV) &
$\Gamma^{(1)}$/$\Gamma^{(0)}$  \\
\hline
1.4&2.14&0.298&176&162&0.919 \\
\hline
1.5&2.28&0.290&151&139&0.918 \\
\hline
1.6&2.42&0.282&131&120&0.918 \\
\hline\hline
\end{tabular}
\caption{The partial decay width with different charm quark mass $m_c$ and 
renormalization scale $\mu=\mu^*$ in BLM scheme. }
\label{table:resultJ}
\end{center}
\end{table}
\begin{table}[htbp]
\begin{center}
\begin{tabular}{|c|c|c|c|c|c|c|}
\hline\hline
$m_b$(GeV) & $\mu^*$(GeV) & $\alpha_s(\mu^*)$ & $\Gamma^{(0)}$(keV) & $\Gamma^{(1)}$(keV) &
$\Gamma^{(1)}$/$\Gamma^{(0)}$  \\
\hline
4.65&6.18&0.204&17.8&18.3&1.03 \\
\hline
4.75&6.29&0.203&17.2&17.7&1.03 \\
\hline
4.85&6.39&0.202&16.6&17.1&1.03 \\
\hline\hline
\end{tabular}
\caption{The partial decay width with different bottom quark mass $m_b$ and
renormalization scale $\mu=\mu^*$ in BLM scheme. }
\label{table:resultU}
\end{center}
\end{table}
It can be
seen that the convergences of the perturbative expansions are all
improved and the K factor is even lower than 1 for the $J/\psi$
production.

The above two schemes give almost the same results for both $J/\psi$ and $\Upsilon$ process.
In the following discussion we will adopt the results from the BLM scheme.

\section{III. Photon and gluon fragmentation processes and the total results}

\begin{figure}
\center{
\includegraphics[scale=0.65]{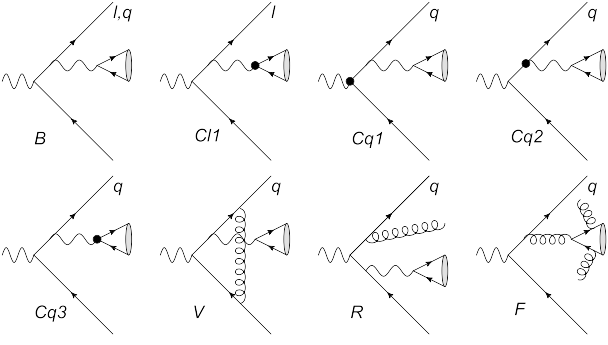}
\caption{\label{fig:qed}The typical Feynman diagrams for the 
fragmentation processes on $Z \to J/\psi+X$. B is the Born diagram.
Cl1, Cq1$\sim$Cq3 are the counterterm diagrams with it's corresponding 
loop diagrams for $J/\psi+l^+l^-+X$ and $J/\psi+q\bar{q}+X$ separately, 
V is the box diagram, R is the diagram for 
the real correction, and F is the gluon fragmentation process.}}
\end{figure}

There are some QED processes which can give contributions comparable
to that of the QCD ones in heavy quarkonium production
\cite{Liu:2003zr}. The contribution from the photon fragmentation processes was investigated 
in reference \cite{Fleming:1993fq} and it gives non-ignorable contribution to the inclusive
$J/\psi$ production in Z boson decay. Therefore, we further
investigate the QCD correction to this photon fragmentation
processes. At leading order, the following processes must be
included,
\begin{eqnarray}\label{singlet}
Z \to J/\psi +l^++l^- \\
Z \to J/\psi +q+\bar{q}.
\end{eqnarray}
Here $l(q)$ is the lepton(quark) and the final results must be summed
over $e$, $\mu$ and $\tau$($u,d,c,s,b$). We only pick out the photon
fragmentation diagrams to calculate. These diagrams form a gauge
invariance subgroup. All the typical Feynman diagrams at LO 
and NLO are shown in Fig.~\ref{fig:qed}.

There are also the  gluon fragmentation processes in CSM,
\begin{eqnarray}\label{singlet}
Z \to J/\psi +q+\bar{q}+g+g.
\end{eqnarray}
Here the $q\bar{q}$ in the final states will be summed over
$u,d,c,s,b$.
Although they are at order $\alpha\alpha_s^4$, the contribution of them
is not too small~\cite{Cheung:1995ka,Baek:1996kq}. The typical 
Feynman diagrams are shown in Fig.~\ref{fig:qed}. 

\begin{figure}
\center{
\includegraphics*[scale=0.4]{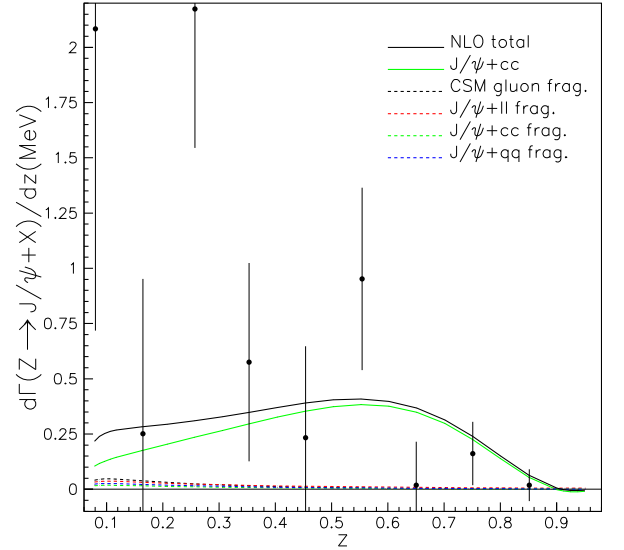}
\caption {\label{fig:tot}The $J/\psi$ energy distribution in $Z\rightarrow
J/\psi+X$ with $m_c=1.4$ GeV, and $\mu=\mu_{BLM}$ for
$J/\psi+c\bar{c}$ and $\mu=2m_c$ for other processes.
}}
\end{figure}

In evaluating these fragmentation processes, we set the renormalization scale
as $2m_c(2m_b)$. The NLO $\alpha_s$ and wave function for quarkonium are also
used. Taking all the above processes in to account, we get the full
results on the partial widths in Table ~\ref{table:resultqed} and the
energy distribution in Fig.~\ref{fig:tot}.

\begin{table}[htbp]
\begin{center}
\begin{tabular}{|c|c|c|c|c|c|}
\hline\hline
$m_c$(GeV) & $\Gamma^{BLM}_{J/\psi+c\bar{c}}$ & $\Gamma_{QCD}^{gluon}$ & $\Gamma_{QED}^{e,\mu,\tau}$ & $\Gamma_{QED}^{u,d,s}$ & $\Gamma_{QED}^{c}$  \\
\hline
1.4&162&9.21&10.5&6.26&4.36 \\
\hline
1.6&120&5.41&8.12&4.91&3.43 \\
\hline\hline
\end{tabular}
\caption{The mass of charm quark is chosen as 1.4 GeV and 1.6 GeV, $\mu=\mu_{BLM}$ for
$J/\psi+c\bar{c}$ and $\mu=2m_c$ for other processes. $\Gamma_{QCD}^{gluon}$ and $\Gamma_{QED}$ 
present the contributions of the photon and gluon fragmentation processes respectively.
(unit of decay widths: KeV)}
\label{table:resultqed}
\end{center}
\end{table}

Combining all the above results together and timing a factor of 1.29
to include the contribution from the $\psi'$ feed-down, we
obtain the branching ratio of $J/\psi$ production in Z decay as
following:
\begin{eqnarray}\label{brtot}
Br_{NLO}^{J/\psi+c\bar{c}+X}=(6.20 \sim 8.38) \times 10^{-5}, \\
Br^{frag. pro.}=(1.13 \sim 1.57) \times 10^{-5},\\
Br^{total}=(7.33 \sim 9.95) \times 10^{-5}.
\end{eqnarray}
Here we give the range of the branching ratio with the charm mass changing
from 1.4 to 1.6 GeV. The total theoretical result is almost the half of the central
value of the experimental measurement in Eq.(\ref{exj}). It is shown in 
Fig.~\ref{fig:tot} that the photon and gluon fragmentation processes
contribute more in the lower energy region and the energy
distribution can not fit the experimental data.

Furthermore, we defined a ratio as
\begin{eqnarray}\label{ratio}
R_{c\bar{c}}=\frac{\Gamma(Z \to J/\psi + c\bar{c}+X)}{\Gamma(Z \to J/\psi + X)}.
\end{eqnarray}
Using the theoretical results obtained in the CSM, the ratio is about 
$R^{CSM}_{c\bar{c}}=0.84~(0.85)$ with $m_c=1.4~(1.6)$ GeV. If we assume that the deriviation 
of the theoretical prediction from the central value of the experimental results is 
from gluon fragmentation processes in the COM that was investigated 
in reference~\cite{Cheung:1995ka,Baek:1996kq}, the ratio can be modified as
\begin{eqnarray}\label{mdfratio}
&R_{c\bar{c}}=\frac{1}{\Gamma_{EX}(Z \to J/\psi + X)}
           \{\Gamma_{CSM}(Z \to J/\psi + c\bar{c}) \nonumber\\
       &+R_{c\bar{c}}^{o}[\Gamma_{EX}(Z \to J/\psi X) 
       -\Gamma_{CSM}(Z \to J/\psi X)]\},
\end{eqnarray}
where $R_{c\bar{c}}^{o}$ from gluon fragmentation processes in the COM is defined as
\begin{eqnarray}\label{rccfrag}
R_{c\bar{c}}^{o}=\frac{\Gamma^{g\to^3S_1(8)}(Z \to c\bar{c}+J/\psi+X)}
          {\sum\limits_q \Gamma^{g\to^3S_1(8)}(Z \to q\bar{q}+J/\psi+X)},
\end{eqnarray}
and the $q\bar{q}$ in the denominator are summed over u, d, c, s, b, 
and $R_{c\bar{c}}^{o}$=0.17 is obtained from reference~\cite{Baek:1996kq}.
Then we obtain $R^{CSM+COM}_{c\bar{c}}=0.49~(0.41)$ for $m_c=1.4~(1.6)$GeV.
The above analysis indicate that $R_{cc}$ can be used to clarify the COM contribution. 

\section{IV. Summary and conclusion}
We have investigated all the processes that give main contributions
to $J/\psi$ inclusive production in Z boson decay in the CSM. 
The results with NLO QCD correction are obtained. For the $Z \to
J/\psi+c\bar{c}$ process, the NLO results only change the leading
order results lightly, and the K factor is 1.13 with $\mu$=2$m_c$ and 
insensitive to the charm quark mass. We also use two methods to 
estimate the dependence of the results on the choice of renormalization scale, 
and these two methods give almost the same partial decay width. 
The K factor even fall to 0.918 by using the BLM scheme. 
We also include 
the contributions of main fragmentation processes. The total branching ratio 
for $Z \to J/\psi+X$ in CSM is $(7.3 \sim 10.0)\times 10^{-5}$, about half of the central value 
of the experimental data 2.1$\times10^{-4}$. 
We define $R_{c\bar{c}}=\frac{\Gamma(Z \to J/\psi  c\bar{c}X)}{\Gamma(Z \to J/\psi  X)}$
and obtain $R_{cc}=0.84$ for only CSM contribution and $R_{cc}=0.49$ for COM and CSM 
contribution together. Then $R_{cc}$ measurement could be used to clarify 
the COM contributions. 
In addition the $J/\psi$ energy
distribution is inconsistent with the experimental data too. But there are large
uncertainties in the experiment results on the inclusive production
of $J/\psi$ in Z decay, not only the total branching ratio but also the
$J/\psi$ energy distribution. Further experimental measurement with
more sample data is needed to clarify the situation. Maybe in the future Z factory
these processes could obtain a detailed investigation. In the calculation, 
the K factor for vector and axial-vector parts of $Z \to
J/\psi+c\bar{c}+X$ are almost same. It may 
indicate that the mechanism of heavy quark fragmentation into quarkonium is dominant 
in this process even at NLO.

\section{Acknowledgments}
We would like to thank Bin Gong and Hong-Fei Zhang for helpful discussion. This
work is supported by the National Natural Science Foundation of
China (No.10475083, 10979056 and 10935012) and by the Chinese Academy of Science under
Project No. INFO-115-B01, and by the China Postdoctoral Science foundation
(20090460525).


\begin{thebibliography}{99}

\bibitem{Bodwin:1994jh}
  G.~T.~Bodwin, E.~Braaten and G.~P.~Lepage,
  Phys.\ Rev.\  D {\bf 51}, 1125 (1995)
  [Erratum-ibid.\  D {\bf 55}, 5853 (1997)]
  [arXiv:hep-ph/9407339].


\bibitem{Einhorn:1975ua}
  M.~B.~Einhorn and S.~D.~Ellis,
  Phys.\ Rev.\  D {\bf 12}, 2007 (1975);
  S.~D.~Ellis, M.~B.~Einhorn and C.~Quigg,
  Phys.\ Rev.\ Lett.\  {\bf 36}, 1263 (1976);
  C.~H.~Chang,
  Nucl.\ Phys.\  B {\bf 172}, 425 (1980);
  E.~L.~Berger and D.~L.~Jones,
  Phys.\ Rev.\  D {\bf 23}, 1521 (1981);
  R.~Baier and R.~Ruckl,
  Nucl.\ Phys.\  B {\bf 201}, 1 (1982).


\bibitem{Brambilla:2004wf}
  N.~Brambilla {\it et al.}  [Quarkonium Working Group],
  arXiv:hep-ph/0412158;
  M.~Kramer,
  Prog.\ Part.\ Nucl.\ Phys.\  {\bf 47}, 141 (2001);
  J.~P.~Lansberg,
  Int.\ J.\ Mod.\ Phys.\  A {\bf 21}, 3857 (2006).


\bibitem{Abe:2001za}
  K.~Abe {\it et al.}  [BELLE Collaboration],
  Phys.\ Rev.\ Lett.\  {\bf 88}, 052001 (2002);
%
  K.~Abe {\it et al.}  [Belle Collaboration],
  Phys.\ Rev.\ Lett.\  {\bf 89}, 142001 (2002);
%
  K.~Abe {\it et al.}  [Belle Collaboration],
  Phys.\ Rev.\  D {\bf 70}, 071102 (2004);
%
  P.~Pakhlov {\it et al.}  [Belle Collaboration],
  Phys.\ Rev.\  D {\bf 79}, 071101 (2009).

\bibitem{Aubert:2005tj}
  B.~Aubert {\it et al.}  [BABAR Collaboration],
  Phys.\ Rev.\  D {\bf 72}, 031101 (2005).


\bibitem{Zhang:2005cha}
  Y.~J.~Zhang, Y.~j.~Gao and K.~T.~Chao,
  Phys.\ Rev.\ Lett.\  {\bf 96}, 092001 (2006);
%
  Y.~J.~Zhang and K.~T.~Chao,
  Phys.\ Rev.\ Lett.\  {\bf 98}, 092003 (2007);
%
  Y.~J.~Zhang, Y.~Q.~Ma and K.~T.~Chao,
  Phys.\ Rev.\  D {\bf 78}, 054006 (2008);
%
  Y.~Q.~Ma, Y.~J.~Zhang and K.~T.~Chao,
  Phys.\ Rev.\ Lett.\  {\bf 102}, 162002 (2009);
%
  B.~Gong and J.~X.~Wang,
  Phys.\ Rev.\  D {\bf 77}, 054028 (2008);
%
  B.~Gong and J.~X.~Wang,
  Phys.\ Rev.\ Lett.\  {\bf 100}, 181803 (2008);
%
  B.~Gong and J.~X.~Wang,
  Phys.\ Rev.\ Lett.\  {\bf 102}, 162003 (2009);
%
  W.~L.~Sang and Y.~Q.~Chen,
  arXiv:0910.4071 [hep-ph];
%
  D.~Li, Z.~G.~He and K.~T.~Chao,
  Phys.\ Rev.\  D {\bf 80}, 114014 (2009);
%
  Y.~J.~Zhang, Y.~Q.~Ma, K.~Wang and K.~T.~Chao,
  Phys.\ Rev.\  D {\bf 81}, 034015 (2010).


\bibitem{Gong:2009ng}
  B.~Gong and J.~X.~Wang,
  Phys.\ Rev.\  D {\bf 80}, 054015 (2009).


\bibitem{Bodwin:2006ke}
  G.~T.~Bodwin, D.~Kang, T.~Kim, J.~Lee and C.~Yu,
  AIP Conf.\ Proc.\  {\bf 892}, 315 (2007);
%
  Z.~G.~He, Y.~Fan and K.~T.~Chao,
  Phys.\ Rev.\  D {\bf 75}, 074011 (2007);
%
  G.~T.~Bodwin, J.~Lee and C.~Yu,
  Phys.\ Rev.\  D {\bf 77}, 094018 (2008);
%
  Z.~G.~He, Y.~Fan and K.~T.~Chao,
  Phys.\ Rev.\  D {\bf 81}, 054036 (2010);
%
  Y.~Jia,
  arXiv:0912.5498 [hep-ph].


\bibitem{Campbell:2007ws}
  J.~M.~Campbell, F.~Maltoni and F.~Tramontano,
  Phys.\ Rev.\ Lett.\  {\bf 98}, 252002 (2007).

\bibitem{Gong:2008sn}
  B.~Gong and J.~X.~Wang,
  Phys.\ Rev.\ Lett.\  {\bf 100}, 232001 (2008);
%
  B.~Gong and J.~X.~Wang,
  Phys.\ Rev.\  D {\bf 78}, 074011 (2008).

\bibitem{Gong:2008ft}
  B.~Gong, X.~Q.~Li and J.~X.~Wang,
  Phys.\ Lett.\  B {\bf 673}, 197 (2009).

\bibitem{Artoisenet:2008fc}
  P.~Artoisenet, J.~M.~Campbell, J.~P.~Lansberg, F.~Maltoni and F.~Tramontano,
  Phys.\ Rev.\ Lett.\  {\bf 101}, 152001 (2008).


\bibitem{Brodsky:2009cf}
  S.~J.~Brodsky and J.~P.~Lansberg,
  Phys.\ Rev.\  D {\bf 81}, 051502 (2010);
%
  J.~P.~Lansberg,
  arXiv:1003.4319 [hep-ph].

\bibitem{Ma:2010vd}
  Y.~Q.~Ma, K.~Wang and K.~T.~Chao,
  arXiv:1002.3987 [hep-ph].


\bibitem{Kramer:1995nb}
  M.~1.~Kramer,
  Nucl.\ Phys.\  B {\bf 459}, 3 (1996).


\bibitem{Artoisenet:2009xh}
  P.~Artoisenet, J.~M.~Campbell, F.~Maltoni and F.~Tramontano,
  Phys.\ Rev.\ Lett.\  {\bf 102}, 142001 (2009);
%
  C.~H.~Chang, R.~Li and J.~X.~Wang,
  Phys.\ Rev.\  D {\bf 80}, 034020 (2009).


\bibitem{Butenschoen:2009zy}
  M.~Butenschoen and B.~A.~Kniehl,
  Phys.\ Rev.\ Lett.\  {\bf 104}, 072001 (2010).


\bibitem{Li:2008ym}
  R.~Li and J.~X.~Wang,
  Phys.\ Lett.\  B {\bf 672}, 51 (2009);
  J.~P.~Lansberg,
  Phys.\ Lett.\  B {\bf 679}, 340 (2009).


\bibitem{He:2009cq}
  Z.~G.~He, R.~Li and J.~X.~Wang,
  arXiv:0904.1477 [hep-ph];
  Z.~G.~He, R.~Li and J.~X.~Wang,
  Phys.\ Rev.\  D {\bf 79}, 094003 (2009).


\bibitem{He:2009by}
  Z.~G.~He and J.~X.~Wang,
  Phys.\ Rev.\  D {\bf 81}, 054030 (2010).


\bibitem{Guberina:1980dc}
  B.~Guberina, J.~H.~Kuhn, R.~D.~Peccei and R.~Ruckl,
  Nucl.\ Phys.\  B {\bf 174}, 317 (1980);
%
  W.~Y.~Keung,
  Phys.\ Rev.\  D {\bf 23}, 2072 (1981);
%
  K.~J.~Abraham,
  Z.\ Phys.\  C {\bf 44}, 467 (1989);
%
  V.~D.~Barger, K.~m.~Cheung and W.~Y.~Keung,
  Phys.\ Rev.\  D {\bf 41}, 1541 (1990);
%
  K.~Hagiwara, A.~D.~Martin and W.~J.~Stirling,
  Phys.\ Lett.\  B {\bf 267}, 527 (1991)
  [Erratum-ibid.\  B {\bf 316}, 631 (1993)];
%
  E.~Braaten, K.~m.~Cheung and T.~C.~Yuan,
  Phys.\ Rev.\  D {\bf 48}, 4230 (1993);
%
  J.~Jalilian-Marian,
  arXiv:hep-ph/9401229;
%
  P.~Ernstrom, L.~Lonnblad and M.~Vanttinen,
  Z.\ Phys.\  C {\bf 76}, 515 (1997);
%
  G.~A.~Schuler,
  Int.\ J.\ Mod.\ Phys.\  A {\bf 12}, 3951 (1997).


\bibitem{Cheung:1995ka}
  K.~m.~Cheung, W.~Y.~Keung and T.~C.~Yuan,
  Phys.\ Rev.\ Lett.\  {\bf 76}, 877 (1996);
%
  P.~L.~Cho,
  Phys.\ Lett.\  B {\bf 368}, 171 (1996);


\bibitem{Baek:1996kq}
  S.~Baek, P.~Ko, J.~Lee and H.~S.~Song,
  Phys.\ Lett.\  B {\bf 389}, 609 (1996);


\bibitem{Fleming:1993fq}
  S.~Fleming,
  Phys.\ Rev.\  D {\bf 48}, 1914 (1993).


\bibitem{Acciarri:1998iy}
  M.~Acciarri {\it et al.}  [L3 Collaboration],
  Phys.\ Lett.\  B {\bf 453}, 94 (1999).

\bibitem{Gregores:1996ek}
  E.~M.~Gregores, F.~Halzen and O.~J.~P.~Eboli,
  Phys.\ Lett.\  B {\bf 395}, 113 (1997).

\bibitem{Boyd:1998km}
  C.~G.~Boyd, A.~K.~Leibovich and I.~Z.~Rothstein,
  Phys.\ Rev.\  D {\bf 59}, 054016 (1999).

\bibitem{Korner:1991sx}
  J.~G.~Korner, D.~Kreimer and K.~Schilcher,
  Z.\ Phys.\  C {\bf 54}, 503 (1992).

\bibitem{Yang:1950rg}
  C.~N.~Yang,
  Phys.\ Rev.\  {\bf 77}, 242 (1950).


\bibitem{Gong:2007db}
  B.~Gong and J.~X.~Wang,
  Phys.\ Rev.\  D {\bf 77}, 054028 (2008).

\bibitem{Wang:2004du}
  J.~X.~Wang,
  Nucl.\ Instrum.\ Meth.\  A {\bf 534}, 241 (2004).

\bibitem{Amsler:2008zz}
  C.~Amsler {\it et al.}  [Particle Data Group],
  Phys.\ Lett.\  B {\bf 667}, 1 (2008).

\bibitem{Brodsky:1982gc}
  S.~J.~Brodsky, G.~P.~Lepage and P.~B.~Mackenzie,
  Phys.\ Rev.\  D {\bf 28}, 228 (1983).

\bibitem{Liu:2003zr}
  K.~Y.~Liu, Z.~G.~He and K.~T.~Chao,
  Phys.\ Rev.\  D {\bf 68}, 031501 (2003).


\end{thebibliography}
\end{document}